# Anomalous Spin Correlations and Mass-Generating Excitonic Instability of Interacting Weyl Fermions

Michihiro Hirata[1,2*], Kyohei Ishikawa[2], Genki Matsuno[3], Akito Kobayashi[3], Kazuya Miyagawa[2], Masafumi Tamura[4], Claude Berthier[5], Kazushi Kanoda[2*]

[1]*Institute for Materials Research, Tohoku University, Aoba-ku, Sendai 980-8577, Japan*

[2]*Deparment of Applied Physics, University of Tokyo, Bunkyo-ku, Tokyo 113-8656, Japan*

[3]*Department of Physics, Nagoya University, Chikusa-ku, Nagoya 464-8602, Japan*

[4]*Department of Physics, Faculty of Science and Technology, Tokyo University of Science, Noda, Chiba 278-8510, Japan*

[5]*Laboratoire National des Champs Magnétique Intenses, LNCMI-CNRS (UPR3228), EMFL, UGA, UPS, and INSA, Boîte Postale 166, 38042 Grenoble Cedex 9, France*

[*]Correspondence to: michihiro_hirata@imr.tohoku.ac.jp and kanoda@ap.t.u-tokyo.ac.jp

**Abstract**: Recent advances in the study of nodal Weyl fermions (WFs), quasi-relativistic massless particles, constitute a novel realm of quantum many-body phenomena. The Coulomb interaction in such systems, having a zero density of states at the Fermi level, is of particular interest, since in contrast to conventional correlated metals, its long-ranged component is unscreened. Here, through nuclear-magnetic-resonance (NMR) measurements, we unveil the exotic spin correlations of two-dimensional WFs in an organic material, causing a divergent increase of the Korringa ratio by a factor of 1000 upon cooling, in striking contrast with conventional metallic behaviors. Combined with model calculations, we show that this divergence stems from the interaction-driven velocity renormalization that almost exclusively suppresses the zero-momentum spin fluctuations. At low temperatures, the NMR rate shows a remarkable increase, which is shown by numerical analyses to correspond to inter-node excitonic fluctuations, precursor of a transition from massless to massive quasiparticles.

The WFs are massless quasiparticles in matter exhibiting a linear energy-momentum dispersion, with low-energy properties governed by the relativistic Dirac-Weyl theory (*1*). They have been discovered in a range of materials in two-dimensional (2D) and three-dimensional (3D) systems, where extensive studies focusing upon their relativistic and topological aspects revealed unconventional charge (*2*) and spin (*3-6*) responses. In contrast to the massive electron systems subjected to short-range electronic correlations, the Coulomb interaction among WFs has a highly unusual characteristic, since its long-range component is unscreened at the band-crossing nodes due to the vanishing density of states (*2*). As a result, anomalous phenomena directly induced by the long-range interactions come along, in which the strength of the interaction is characterized by a dimensionless coupling constant $\alpha$, given by the ratio of the Coulomb potential to the electronic kinetic energy (*1*, *2*). For instance, in a weak coupling regime, recent studies have found an upward renormalization of the electron velocity in the 2D system graphene (*1*, *2*, *7*), in marked contrast to its suppression in conventional correlated materials. In strong coupling, theoretical studies have predicted an even more exotic phenomenon: an excitonic mass gap opening, which originates from the incipient instability of massless fermions, as first discussed in high-energy physics (*8*, *9*) and more recently in the context of condensed matter (*10-12*). In spite of a great deal of theoretical advances, however, the size of $\alpha$ proves to be rather small in a range of materials (*1*, *2*); consequently, experimental characterization of WFs has remained largely limited especially under strong coupling, and the excitonic instability is yet to be investigated.

Here, by combining NMR experiments and model calculations, we demonstrate the realization of a strongly-coupled 2D WF system in the organic salt $\alpha$-(BEDT-TTF)$_2$I$_3$ ($\alpha$-I$_3$) and report the observation of unconventional spin dynamics that are governed by the interaction-induced velocity renormalization and mass-generating excitonic instability. We find that the temperature ($T$) driven velocity renormalization develops on cooling and suppresses only the zero-momentum mode of the spin susceptibility. This makes a contrasting impact on the $T$ dependence of the $^{13}$C Knight shift $K$ and the spin-lattice relaxation rate $1/T_1T$, leading to a divergent enhancement of the Korringa ratio ($\propto 1/T_1TK^2$) by orders of magnitude. Although the renormalization reduces the size of $\alpha$, known as the running coupling constant (*1*, *2*), we show that it remains considerably large at low temperatures in $\alpha$-I$_3$, giving rise to anomalous spin fluctuations that signify a precursor of mass-generating excitonic instability.

The organic charge-transfer salt $\alpha$-I$_3$ is a layered material comprised of BEDT-TTF (ET) conducting and I$_3$ insulating layers (Fig. 1A). A pair of Weyl nodes formed by ET molecular orbitals appear at $\mathbf{k}_0$ and $-\mathbf{k}_0$ in the 2D momentum space upon destabilizing the nonmagnetic charge-ordered phase, induced by the short-range repulsions (*13-15*), with a hydrostatic pressure of $\geq P_C = 1.2$ GPa (Fig. 1B). The 2D WF phase having the titled-Dirac-cone dispersion (Fig. 3C) emerges in the conducting layers at low temperature (*13, 16-18*), in which the Weyl nodes are fixed at the Fermi energy $E_F$ by virtue of the 3/4-filled nature of the energy band (*16, 17*). The presence of WFs competing with charge order suggests a strong influence of the electron-electron interaction on the nature of WFs (*4, 6, 15, 19-21*). In fact, recent Knight shift measurements above $P_C$ (*3*) found a logarithmic velocity enhancement due to the long-range part of the Coulomb interaction as well as a ferrimagnetic spin polarization by the short-range repulsions (*6, 21*). These findings make $\alpha$-I$_3$ an ideal playground for investigating interaction effects of WFs. In addition, a conventional 2D metal-like regime shows up at high temperatures (*3*), reflecting the density of states (DOS) that possesses a flat profile approximately above $|E_W - E_F|$ ~12meV (~150 K) (insets of Fig. 2A) (*16, 22*). Varying the experimental energy scale (i.e., temperature), one can thus explore the electronic properties of WFs at low $T$ as compared to the 2D metal-like state at high $T$, by making use of the conventional knowledge of interaction studies established in ordinary massive electron systems.

For obtaining deeper knowledge about the electron-electron interactions, $1/T_1T$ proves to be a powerful means (*23-25*) which probes the wavenumber $\mathbf{q}$ average of the dynamic spin susceptibility $\chi(\mathbf{q}, \omega)$ ($\omega$: frequency of MHz). Figure 2A presents the temperature dependence of $1/T_1T$ and the squared Knight shift $K^2$ in the WF state of $\alpha$-I$_3$ (at 2.3 GPa), measured at the $^{13}$C nuclei in the molecule A. (A magnetic field of 6 T is applied parallel to the conducting layers for NMR measurements; for details, see Ref. *26*.) While a *T*-independent nature appears above 150 K due to the flat DOS above $|E_W - E_F|$, both $1/T_1T$ and $K^2$ show below approximately 150 K a drastic decrease upon cooling reflecting the vanishing DOS around the node on $E_F$ (insets of Fig. 2A). In conventional massive electron systems, the ratio of $1/T_1T$ to the square of *K*, $\mathcal{K} = 1/(T_1TS_0\beta K^2)$, named the Korringa ratio, measures the strength of the short-range electron correlations (*24, 25*), where $S_0 = (4\pi k_B/\hbar)(\gamma_n/\gamma_e)^2$ with the electron (nuclear) gyromagnetic ratio $\gamma_e$ ($\gamma_n$) and $\beta$ is a form factor standing for the anisotropy of the nuclear hyperfine interaction (*26*). The value of $\mathcal{K}$ is of the order of unity and does not considerably vary with *T* for weak correlations (the Korringa law), whereas a sizable deviation of $\mathcal{K}$ from unity occurs for noticeable correlations, where $\mathcal{K} > (<) 1$ stands for enhanced antiferromagnetic (ferromagnetic) spin fluctuations (*24, 25*). Above 150 K, we observe that the Korringa law holds and $\mathcal{K}$ is ~3 (Fig. 2B), such as have been observed in typical 2D organic metals $\theta$-(ET)$_2$I$_3$ (*27*) and $\kappa$-(ET)$_2$Cu(NCS)$_2$ (*28*), corroborating the 2D metallic picture with moderate short-range repulsions.

On the contrary, a spectacular breakdown of the Korringa law sets in below 150 K, and an orders-of-magnitude increase of $\mathcal{K}$ appears with decreasing temperature, leading to $\mathcal{K}$ ~$10^3$ at 10 K (Fig. 2B). The remarkable increase of $\mathcal{K}$ directly originates from the distinct *T* dependence of $K^2$ and $1/T_1T$ (Fig. 2A), namely, $K^2$ drops much faster than $1/T_1T$ on cooling. This is in clear contrast with the non-interacting WFs that satisfy the Korringa law and produce $\mathcal{K} \approx 1.71$ in the DOS approximation (*29*). In conventional correlated systems, the increase of $\mathcal{K}$ is attributed to the growing finite-**q** antiferromagnetic fluctuations that push up the **q** average of the dynamic susceptibility $\chi(\mathbf{q}, \omega)$ ($\propto 1/T_1T$) while the static uniform susceptibility $\chi(\mathbf{q} \to 0, 0)$ ($\propto K$) is kept intact; the enhancement of $\mathcal{K}$ is, for example, 10 or less in highly correlated metals near a Mott transition (*23*). The extraordinarily large $\mathcal{K} \sim 10^3$ and its prominent temperature evolution, therefore, point to a novel spin-correlation mechanism which is intrinsic to massless fermion systems.

For 2D WFs, one has to recall that the conduction and valence bands touch at $E_F$ and constitute two Weyl nodes at $\pm\mathbf{k}_0$ in momentum space, around which the Coulomb interaction remains long ranged owing to the lack of conventional screening (*1, 2*). The electronic excitations at low temperature exclusively appear around these nodes, which can be sorted into two processes that are characterized by a contrasting momentum transfer $\hbar\mathbf{q}$, $C_1$ ($\mathbf{q} \sim 0$) and $C_2$ ($\mathbf{q} \sim 2\mathbf{k}_0$) in Fig. 3C. The NMR relaxation rate $1/T_1T$ probes the sum of $C_1$ and $C_2$, whereas the shift $K$ sees only $C_1$ (in particular $\mathbf{q} = 0$). Therefore, the different profile of the traces in Fig. 2A suggests that the Coulomb interaction provides contrary influence upon two processes.

To evaluate the impacts of the long-range component of the Coulomb interaction on $C_1$ and $C_2$, we have performed a renormalization group (RG) calculation at the one-loop level for dealing with the self-energy correction, based on the leading-order large-*N* approximation (*3*) and the tilted Weyl Hamiltonian that describes WFs in $\alpha$-I$_3$ (*16-18*) (for details, see Ref. *26*). The bare Coulomb coupling constant, $\alpha \approx e^2/\varepsilon\hbar v$, is estimated to be $\approx 8.4$ by using $\varepsilon \approx 30$ and $v = 2.4\times10^4$ m s$^{-1}$ as determined from fitting $\chi(0, 0)$ (*3*), where $\varepsilon$ is the permittivity and $v$ is the electron velocity (*26*). As shown in Figs. 3, A and B, the RG calculation can properly trace the observed excess suppression of $K^2$ with respect to $1/T_1T$ and hence the divergent increase of $\mathcal{K}$ towards lower temperature. The contrasting temperature dependence of $K^2$ and $1/T_1T$ can be accounted for by the $T$-driven RG flow of the coupling constant (fig. S1) and the resultant upward velocity renormalization, which only suppresses the $\mathbf{q} = 0$ response. The $\mathbf{q} = 0$ static spin susceptibility ($K$), probing $C_1$, is directly affected by the renormalization (Fig. 3E), leading to a continuous drop of $K^2$ on cooling (*3*, *6*). On the contrary, the $\mathbf{q}$-summed $\chi(\mathbf{q}, \omega)$ ($1/T_1T$), probing the sum of $C_1$ and $C_2$, is rather enhanced over the non-interacting value; the $C_1$ process ($\mathbf{q} \sim 0$) dies off on cooling due to the renormalization, whereas the $C_2$ process ($\mathbf{q} \sim 2\mathbf{k}_0$) does not and becomes prevailing at low temperature (Fig. 3D), causing a levelling-off of $1/T_1T$ as presented in Fig. 3A. Furthermore, we find that the tilt of the cone does not affect the results (Fig. 3B). Thus, these calculations demonstrate that the divergent increase of $\mathcal{K}$ is a ubiquitous hallmark of *general* 2D WFs, having either tilted or vertical cones, which is directly promoted by the Coulomb interaction-driven velocity renormalization.

Remarkably, we additionally observe that $1/T_1T$ shows a sudden upturn below 3 K and increases by a factor of 2 towards 1.7 K (Fig. 4A). Relaxations caused by a residual finite DOS at $E_F$ (*29*) or the $C_2$-dominated excitations as presented in Fig. 3A (at low $T$) cannot explain this result, since they produce a levelling-off but not an upturn. Given the sharp nature of the upturn, it is most conceivable that a slow spin dynamics starts to develop at low energies. It is worthwhile to mention that, though the $T$-driven RG flow (fig. S1) reduces the Coulomb coupling $\alpha$ towards lower temperature, its value remains rather large in the present temperature range: $\alpha \approx 2.0$ at 5 K. Considering this sizeable value, the upturn of $1/T_1T$ suggests emergent spin fluctuations associated with the incipient instability of gapless fermions that is driven by the Coulomb interaction.

In the presence of the Coulomb force that preserves its long-ranged nature, theoretical studies in 2D WF systems have revealed an electronic instability favoring excitonic pair formations (*30*). Above a critical value for the Coulomb coupling $\alpha$ ($> \alpha_\mathrm{C} \sim 1$), this leads to an excitonic transition accompanied by mass acquisition (*2, 10, 11*), akin to the chiral symmetry breaking that has been intensively studied in the relativistic high-energy theory (*8, 9*). To examine the influence of this instability, we calculated the dynamic spin susceptibility $\chi(\mathbf{q}, \omega)$ based on the ladder approximation (fig. S2) together with the tilted Weyl Hamiltonian and evaluated the excitonic gap function at a mean-field level, taking properly into account the velocity renormalization (*26*). The excitonic pairings that cause a dynamic mass in 2D WFs can be categorized in terms of time-reversal and node-pseudospin symmetries as well as the total momentum (*11*). Among eight of the possible pairings, we confirmed that the spin-triplet even-parity pairing can produce the upturn of $1/T_1T$ as a clear manifestation of the growing excitonic spin fluctuations (Fig. 4B). The fluctuations set in at a similar temperature to the observation (Fig. 4A) when one assumes a coupling constant of $\alpha = 2$, the size of which well agrees with the RG flow mentioned above (fig. S1).

Furthermore, the calculation shows that the upturn is mainly induced by the inter-node transverse excitonic spin fluctuations (fig. S3), reflecting the fact that the $C_2$ process has a perfect nesting while the $C_1$ process does not for the particle-hole channel in a tilted-cone system, as the cones are tilted in opposite directions in momentum space (see Fig. 3C). This is in good agreement with our argument that the $C_2$ process dominates the low-energy excitations at low temperature (Fig. 3, D and E). Therefore, the upturn of $1/T_1T$ can be perceived as an indication of the precursor to the inter-node excitonic condensation.

Khveshchenko (*11*) has recently estimated the ratio of the mean-field critical temperature $T_C$ to the excitonic gap $\Delta$ for 2D WFs, and by omitting the velocity renormalization, arrived on the expressions

$$\frac{T_C}{\Delta} \approx \frac{1}{|\ln(1 - \tilde{\alpha}_C/\tilde{\alpha})|}, \tag{1}$$

$$\Delta = v\Lambda\exp\left(-\frac{2\pi - 4\arctan\sqrt{2\tilde{\alpha} - 1}}{\sqrt{2\tilde{\alpha} - 1}}\right), \tag{2}$$

where $\tilde{\alpha} = \alpha/(1 + N\pi\alpha/8\sqrt{2})$ with $N = 4$, coming from the two spin projections and two nodes, $\tilde{\alpha}_C = 0.5$ and $\Lambda$ is a momentum cutoff of the inverse lattice constant. For $\alpha$-I$_3$ having a large bare Coulomb coupling ($\alpha \approx 8.4$), the estimated critical temperature and the gap are rather large, $T_C \sim \Delta \sim 10$ K, and are reduced to ~ 40 mK if the RG flow is taken into account ($\alpha \approx 2.0 > \alpha_C$ at the lowest $T$) (*26*). This is consistent with our observation where the incipient excitonic fluctuations appear to develop from higher temperatures towards $T_C$. Furthermore, previous reports of an in-plane insulating behavior in the pressurized $\alpha$-I$_3$ at similar temperatures (*15, 31*) may provide another sign of the precursor of condensation (*32*), in line with our results.

An interesting question is whether the spin transverse fluctuations detected by the NMR rate here would be a common characteristic of interacting WFs. Our calculation indicates that the excitonic instability occurs more directly in the transverse channel than in the longitudinal channel (fig. S3). Given that the tilted cone of $\alpha$-I$_3$ turns out to be rather isotropic at low energy by the velocity renormalization (fig. S1) (*3, 26*), the NMR rate of isotropic WFs, such as those in graphene, would also produce an upturn due to the spin transverse instability.

The experimental and theoretical studies presented here thus cast light on the unconventional dynamics of 2D WFs. We found that the long-range component of the Coulomb interaction, unscreened around the Weyl nodes at $E_F$, not only produces anomalous spin correlations associated with the velocity renormalization but also induces incipient mass-acquisition instability of massless fermions. This work paves the avenue for explorations of the rich physics of "strongly interacting WFs" in solids (*2*), which has, so far, been mainly discussed in a theoretical basis.

**Acknowledgements:**

We thank K. Nomura for critical discussions and reading of the manuscript; D. Basko for technical help with RG calculations; M. O. Goerbig for support and discussions; and J. S. Kinyon for fruitful discussions. We also thank discussions with many colleagues, including H. Fukuyama, N. Nagaosa, H. Isobe, H. Kohno, Y. Suzumura, M. Ogata, T. Osada, C. Hotta, H. Yasuoka, D. Liu, H. Nojiri, T. Kihara, M. Potemski, M.-H. Julien, H. Mayaffre and M. Horvatić. This work is supported by MEXT/JSPS KAKENHI (Grant Noes 20110002, 21110519, 24654101, 25220709, 15K05166 and 15H02108), JSPS Postdoctoral Fellowship for Research Abroad (Grant No. 66, 2013) and MEXT Global COE Program at University of Tokyo (Global Center of Excellence for the Physical Sciences Frontier; Grant No. G04). We also acknowledge support from the Kurata Memorial Hitachi Science and Technology Foundation.

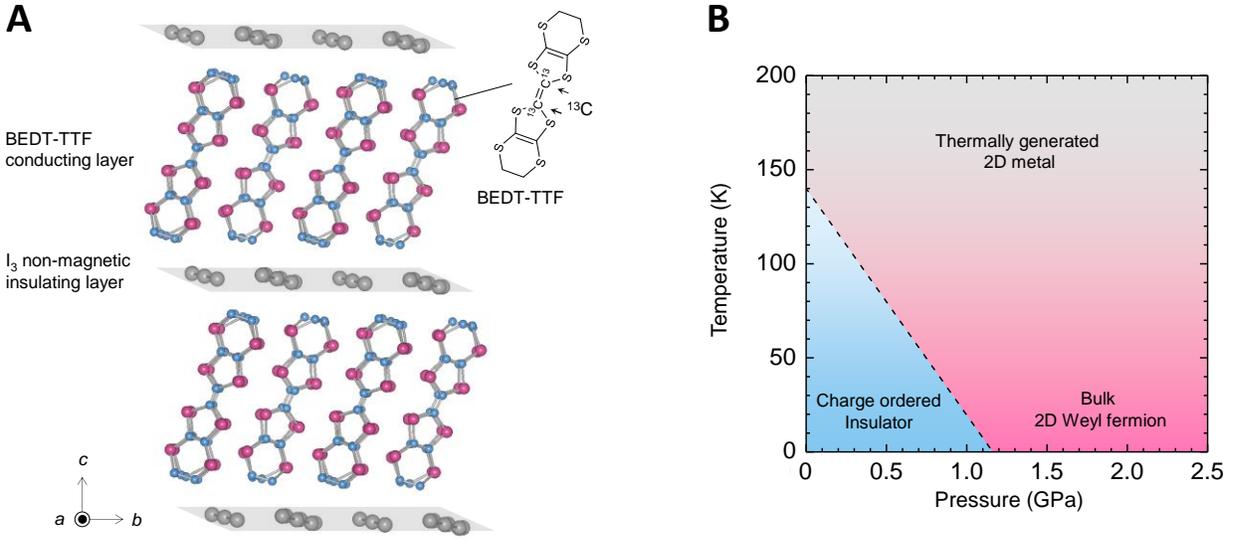

**Fig. 1. Crystal structure and phase diagram of α-(BEDT-TTF)$_2$I$_3$.** (**A**) Side view of the crystal structure. The conducting BEDT-TTF layers are separated by non-magnetic insulating layers of I$_3$. (Inset) BEDT-TTF molecular structure with selectively introduced $^{13}$C isotopes indicated by arrows. (**B**) Pressure-temperature phase diagram of α-(BEDT-TTF)$_2$I$_3$ taken from Ref. *3*. The dashed line indicates the first-order transition line.

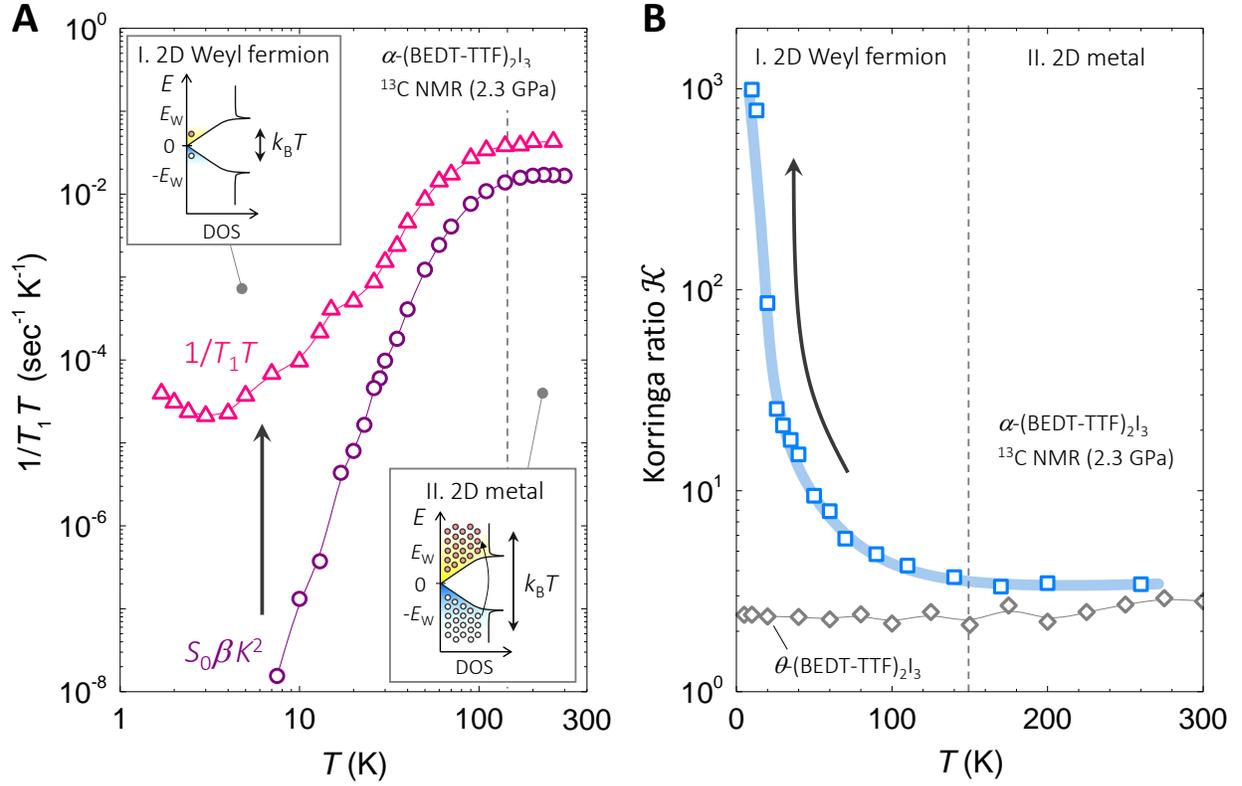

**Fig. 2. Anomalous spin correlation of 2D WFs in $\alpha$-I$_3$.** (**A**) Temperature dependence of the $^{13}$C-NMR spin-lattice relaxation rate $1/T_1T$ (triangles) and the Knight shift squared $K^2$ (circles) measured at a pressure of 2.3 GPa and a magnetic field of 6 T, using the $^{13}$C nuclei at the center of BEDT-TTF molecules (inset of Fig. 1). (Insets) The DOS profile near $E_F$ (= 0) with thermally generated electron-hole pairs (circles) indicated for low (I) and high (II) temperatures, respectively. The DOS is linear up to $|E_W - E_F|$ and levels off above it. (**B**) Temperature dependence of the $^{13}$C-NMR Korringa ratio $\mathcal{K}$ (squares). The results for the 2D organic metal $\theta$-(BEDT-TTF)$_2$I$_3$ (diamonds) is plotted as a reference (*27*) (for details, see the text and Ref. *26*).

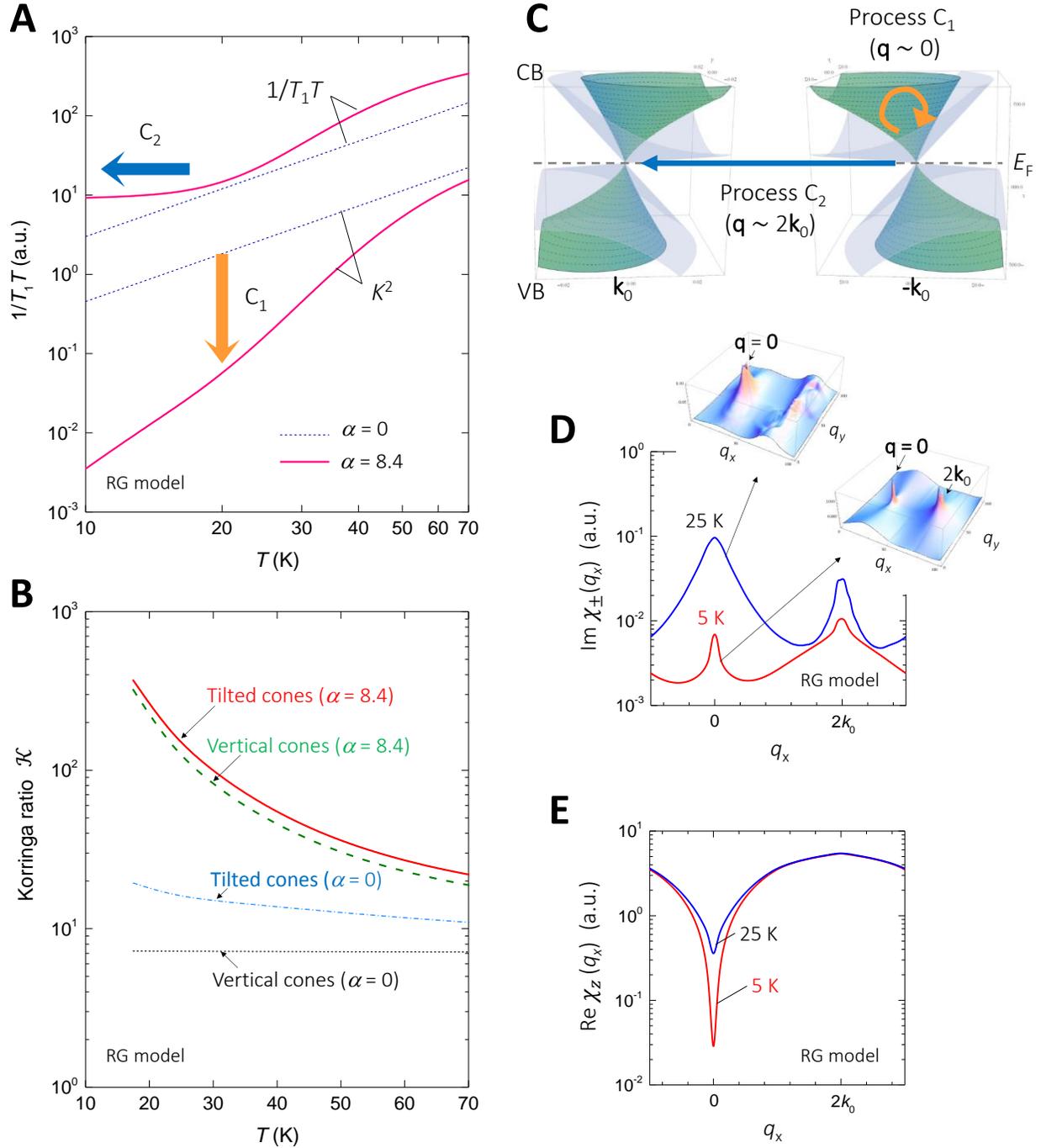

**Fig. 3. RG simulations.** (**A**) Calculated temperature dependence of $1/T_1T$ and $K^2$ for double tilted Weyl cones with the velocity renormalization due to the self-energy correction considered, using the RG technique within the leading-order large-$N$ expansion (*3*). The Coulomb coupling is chosen as $\alpha = 0$ (dashed) and 8.4 (solid) determined from fitting the $K$ data (*3*) (see Ref. *26*). (**B**) Calculated $\mathcal{K}$ as a function of temperature for double cones with and without the tilt and self-energy correction: black dotted, vertical cones ($\alpha = 0$); blue dash-dotted, tilted cones ($\alpha = 0$); green dashed, vertical cones ($\alpha = 8.4$); red solid, tilted cones ($\alpha = 8.4$). (**C**) Double tilted Weyl cones in *a*-I$_3$ for $\alpha = 0$ (grey cones) and 8.4 (green reshaped cones). Two Weyl nodes appear at $\pm\mathbf{k}_0$ where the conduction band (CB) and the valence band (VB) touch at $E_\mathrm{F}$. Relevant excitation processes at low temperature (C$_1$ and C$_2$) are indicated by arrows. (**D, E**) Calculated wavevector dependence of the transverse ($\chi_\pm$) and longitudinal ($\chi_z$) spin susceptibilities for double tilted cones with the self-energy correction ($\alpha = 8.4$). The direction of $q_x$ is set along the line connecting two nodes. The calculated profiles of Im$\chi_\pm$ (**d**) and Re$\chi_z$ (**e**) at 25 K (blue) and 5 K (red) are presented. (Inset of **D**): corresponding 3D plot of Im$\chi_\pm$ on the $q_x$-$q_y$ plane. Note that the $q_x = 0$ term in Re$\chi_z$ corresponds to $K$, while the **q**-summation of Im$\chi_\pm$ amounts to $1/T_1T$. a.u. stands for the arbitrary units.

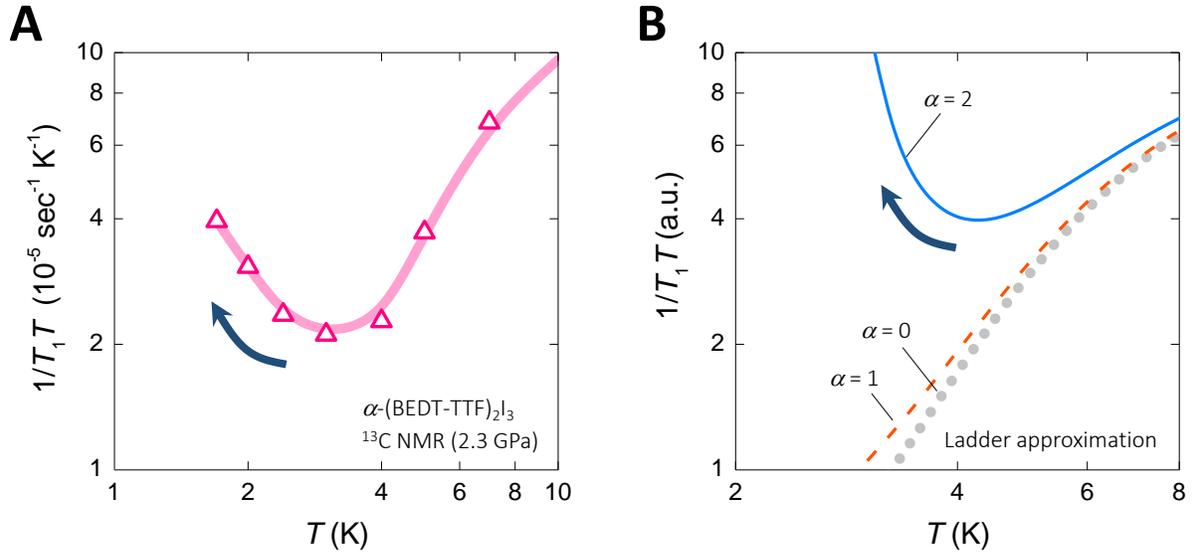

**Fig. 4. Incipient excitonic fluctuations.** (**A**) The low-temperature close up of $1/T_1T$ in Fig. 2A. (**B**) Calculated temperature dependence of $1/T_1T$ based on the ladder approximation at a linearized level for the gap with the velocity renormalization taken into account. The Coulomb coupling of $\alpha = 0$ (grey dotted), 1 (orange dashed) and 2 (blue solid) is used in the calculation (*26*).

**Materials and Methods:**

**Experimental backgrounds.** High-quality single crystals of $\alpha$-(BEDT-TTF)$_2$I$_3$ ($\alpha$-I$_3$) were prepared using conventional methods. Initially, 99% of the double-bond connected central carbon atoms in BEDT-TTF molecules were selectively replaced by carbon-13 ($^{13}$C) isotopes with a nuclear spin-1/2 (upper inset of Fig. 1A). Single crystals were synthesized from the $^{13}$C-enriched BEDT-TTF molecules using electrochemical techniques (the same crystals are used as in Ref. *3*.). A hydrostatic pressure of 2.3 GPa was applied to the sample using a BeCu/NiCrAl clamp-type pressure cell (C&T Factory) and the pressure medium Daphne 7373 (Idemitsu). To prevent solidification of the oil during pressurization, thermal treatments described in Ref. *3* are performed. A static magnetic field $H$ of 6 T was applied parallel to the 2D conducting plane ($H\|ab$). NMR measurements at $^{13}$C nuclei were performed using a commercially available homodyne spectrometer by means of the standard spin-echo technique, with the echo signals recorded at a fixed radiofrequency and converted to NMR spectra via Fourier transformation. The $^{13}$C spin-lattice relaxation time $T_1$ were obtained from the standard single-exponential fits to the recovery of the nuclear magnetization after saturation.

The observed $^{13}$C-NMR spectra are composed of eight lines, the assignment of which can be properly done from the knowledge of the crystal structure as described elsewhere (*3, 33, 34*). The eight lines can be assigned to three non-equivalent molecular sites in the 2D unit cell, often distinguished as A (= A'), B and C (*35*). The total NMR shift, *S*, for a given molecular site is defined as the center-of-mass position of the $^{13}$C lines arising from the corresponding molecule, which is a sum of the spin shift (Knight shift) $K$ and the temperature-independent chemical shift $\sigma$ (i.e., $S = K + \sigma$) with little contributions from the orbital term (*36*). The chemical shift $\sigma$ is determined from the field-angle dependence of the total shift at 3 K where the spin shift is expected to be vanishingly small as we discussed previously (see Ref. *3*), and thus one can assume $S(3\,\mathrm{K}) \approx \sigma(3\,\mathrm{K})$. Subtracting $\sigma$, the total shift $S$ is eventually converted to the Knight shift $K$. The spin-lattice relaxation rate divided by temperature, $1/T_1T$, was measured for each molecule at respective lines, which is related to the imaginary part of the transverse spin susceptibility $\chi^{\perp}$ as follows (*24, 25*)

$$\frac{1}{T_1 T} = \frac{2\gamma_n^2 k_B (\bar{A}_\perp)^2}{g^2 \mu_B^2} \sum_{\mathbf{q}} \frac{\mathrm{Im}\chi^{\perp}(\mathbf{q},\omega_0)}{\omega_0}, \quad (3)$$

where $\mathbf{q}$ is the wavenumber vector, $\gamma_n$ is the nuclear gyromagnetic ratio, $\bar{A}_\perp$ is the transverse component of the mean hyperfine coupling tensor averaged for the two central $^{13}$C nuclei in a molecule (see the inset of Fig. 1A) and $\omega_0$ is the NMR resonance frequency ($\approx 64$ MHz for the $^{13}$C nuclei at 6 T ). Note that the difference in the value of $T_1$ at the two $^{13}$C sites in the center of a molecule is negligibly small and is therefore omitted here.

The impact of the long-range Coulomb interaction is evaluated through comparing the values of $K$ and $1/T_1T$. These quantities are measured at the $^{13}$C NMR lines arising from the molecule A (= A'), where it has been shown that the excitations around a Weyl nodes on the Fermi energy $E_F$ can be most uniformly probed in momentum space, as we discussed previously (*3, 37*). Since the Knight shift in this compound has a large anisotropy due to the anisotropic hyperfine interaction at the $^{13}$C nuclei (*34, 38, 39*), all measurements are performed in a magnetic field applied 60° off the crystalline *a* axis, where the modulus of the Knight shift at the molecule A (= A') becomes maximum. (Note that this configuration is identical to one of those used in Ref. *3*). Hereafter, we omit the site index for simplicity. When the index is needed, it will be specifically mentioned.

In itinerant electron systems, $K$ and $1/T_1T$ satisfy the so-called Korringa relation (*24, 25, 34, 36, 40*), which is given by the following expression

$$\frac{1}{T_1 T K^2} = \frac{4\pi k_B}{\hbar} \left(\frac{\gamma_n}{\gamma_e}\right)^2 \beta \mathcal{K}, \tag{4}$$

where $\gamma_e$ is the electron gyromagnetic ratio and $\mathcal{K}$ is the Korringa ratio, the size of which provides detailed knowledge about spin correlations of conducting electrons. Here, $\beta$ is the form factor describing the anisotropy of the hyperfine coupling (the ratio of the transverse component $\bar{A}_\perp$ to the longitudinal one $\bar{A}_\parallel$) (*34, 36*). For the evaluation of $\mathcal{K}$, we used the principal values of the hyperfine coupling tensor determined at ambient pressure around room temperature (*34*), as the pressure and temperature dependence of this tensor is negligibly small (*3*), which leads to $\beta \sim 0.46$ for the current field geometry, by using the X-ray crystal structure data determined at close pressure (*41*).

**Renormalization-group calculation based on the tilted Weyl Hamiltonian.** Reflecting the very low-symmetric crystal in $\alpha$-I$_3$ containing only the inversion symmetry (*35, 41, 42*), the 2D Weyl nodes with a titled cone dispersion appear at $\mathbf{k}_0$ and $-\mathbf{k}_0$ in the first Brillouin zone due to accidental degeneracy (*43, 44*) above a threshold hydrostatic pressure ($P$) of $P \geq P_C = 1.2$ GPa (*16-18, 22, 45-47*). Due to the 3/4-filled nature of the electronic band, the nodes are anchored at $E_F$ and are shown to be robust against perturbations such as modulations of transfer integrals in a finite range (*16, 37, 48-53*). The low-energy Weyl fermion (WF) state is indeed confirmed by a range of experiments under pressures such as transport (*15, 54-62*), calorimetric (*63, 64*), optical (*65, 66*) and nuclear magnetic resonance (*3, 33, 39*) measurements. The effective model for the 2D WFs is described by the tilted Weyl Hamiltonian (*16-18*) within the Luttinger-Kohn (LK) representation (*67*) based on the molecular orbitals. In the presence of an in-plane magnetic field $H$, the Hamiltonian is given by

$$H_0 = \hbar(\mathbf{w} \cdot \mathbf{k}\hat{\sigma}_0 \otimes \hat{\tau}_z + v_x k_x \hat{\sigma}_x \otimes \hat{\tau}_z + v_y k_y \hat{\sigma}_y \otimes \hat{\tau}_0) \otimes \hat{s}_0 - \mu_B H \hat{\sigma}_0 \otimes \hat{\tau}_0 \otimes \hat{s}_z \tag{5}$$

where $\mathbf{w} = (w_x, w_y)$ and $\mathbf{v} = (v_x, v_y)$ are velocities describing the tilt and the anisotropy of the 2D Weyl cones, respectively; $\mathbf{k} = (k_x, k_y)$ is the wavenumber vector measured from the node $\mathbf{k}_0$; $H$ is the in-plane magnetic field applied parallel to the 2D plane; and $\hat{\sigma}_i$, $\hat{\tau}_j$ and $\hat{s}_t$ are the Pauli matrices representing the LK pseudospin, node-pseudospin and real spin, respectively, with the three indices taking one of the four possible values ($i, j, t = 0, x, y, z$). The index 0 stands for 2×2 unit matrices. Note that we don't distinguish the two nodes and will concentrate on the one at $\mathbf{k}_0$ in this subsection, whereas we shall specifically distinguish them in the subsequent subsection to take into account the inter-node scattering processes between $\mathbf{k}_0$ and $-\mathbf{k}_0$.

For evaluating the long-range Coulomb interaction effects around a node, we have considered the self-energy correction within the frame of the one-loop order renormalization-group (RG) calculation in the leading order in $1/N$ ($N \gg 1$) with $N = 4$ being the number of fermion species (two spin projections and two nodes) (*3*). For the present work, we used the resultant upward flow of the velocity $\mathbf{v}$ (as $\mathbf{w}$ does not flow at the one-loop level) due to the running coupling constant (*2, 6, 68-70*) that is determined from fitting the temperature dependence of the uniform spin susceptibility (Knight shift) (see Ref. *3* and the supplementary materials therein for details of the fitting analyses). The dimensionless Coulomb coupling, which is the ratio of the Coulomb potential to the electron kinetic energy, is given by $\alpha = e^2 / \left( \varepsilon \hbar \sqrt{v_x^2 \sin^2 \varphi + v_y^2 \cos^2 \varphi} \right)$ for the anisotropic 2D WFs (*3*), where $\varepsilon$ is the dielectric constant and $\varphi$ is an angle measured around the node at $\mathbf{k}_0$. As a consequence of the *T*-driven velocity renormalization, the Coulomb coupling $\alpha$ flows to a smaller value (*6, 17, 69, 70*), causing a remarkable reshaping of tilted Weyl cones as depicted in Fig. 3C (*3*).

Concerning the interaction, it is worthwhile to mention that the Coulomb interaction would be quite large in $\alpha$-I$_3$ as the WF state emerges next to a charge-ordered insulator, stabilized by the short-range repulsions (*20, 48, 71-73*) on the *P-T* phase diagram (Fig. 1B) (*15, 56, 66, 74, 75*). Indeed, the bare Coulomb coupling constant, approximated as $\alpha \approx e^2 / \varepsilon \hbar v$ due to the small anisotropy ($v_x \approx v_y \equiv v$) (*3, 37*), is estimated to be $\alpha \approx 8.4$ at the momentum cutoff $q = \Lambda$, (circular around the node), by employing $\varepsilon \approx 30$ and $v = 2.4 \times 10^4$ m s$^{-1}$ determined from the fitting (*3*). We have calculated the flow of $\alpha$ as a function of energy measured from $E_F$ for the gentle and steep slopes of the 2D tilted Weyl cone, as depicted in Fig. S1. The size of $\alpha$ has distinct values in the two slopes at higher energies, while the difference becomes smaller towards lower energy and is eventually negligibly small at the low-energy limit. This is equivalent to say that the tilted cone becomes more isotropic at low energy due to the flow-induced cone reshaping as discussed previously (*3*).

**Ladder approximation for the double-cone model of 2D tilted WFs.** To deal with interacting 2D WFs ($N = 4$) and scatterings between the two nodes, we describe the WF states with an 8-component creation operator $\Psi_{\mathbf{k}}^{\dagger} = (c_{\mathbf{k},s,\eta}^{\nu})^{\dagger}$, where the superscript, the second and third indices of the subscript stand for the two LK bases ($\nu = a, b$) (67), spin ($s = 1(\uparrow), -1(\downarrow)$) and node ($\eta = 1(R), -1(L)$), respectively. Note that the wavenumber vector is given around a node by $\mathbf{k} = \tilde{\mathbf{k}} \mp \tilde{\mathbf{k}}_0$, where $\tilde{\mathbf{k}}$ and $\tilde{\mathbf{k}}_0$ are defined in the first Brillouin zone, and the two Weyl nodes locate at $\pm\tilde{\mathbf{k}}_0$. The effective 8×8 Hamiltonian describing interacting 2D WFs then reads

$$H_{\text{eff}} = \sum_{\mathbf{k}} \Psi_{\mathbf{k}}^{\dagger} H_0 \Psi_{\mathbf{k}} + \sum_{\mathbf{q}} V_0(\mathbf{q})\rho(\mathbf{q})\rho(-\mathbf{q}), \qquad (6)$$

where $V_0(\mathbf{q}) = 2\pi e^2/\epsilon|\mathbf{q}|$ is the long-range Coulomb potential and $\rho(\mathbf{q}) = \sum_{\mathbf{k}} \sum_{\nu,s,\eta} c_{\mathbf{k},s,\eta}^{\nu\dagger} c_{\mathbf{k}+\mathbf{q},s,\eta}^{\nu}$ is the density operator. In the following, we omit the backscattering and Umklapp processes, assuming that both $\mathbf{k}$ and $\mathbf{k} \pm \mathbf{q}$ are restricted to the vicinity of the same node. Hereafter, we thus redefine $\mathbf{q}$ such that it is a wavenumber vector much smaller compared to $2\tilde{\mathbf{k}}_0$.

Remarkably, the spin-lattice relaxation rate at the $l$ th molecular site (or the $l$ th orbital in general), $(1/T_1T)_l$, includes both the intra-node, $\eta = \eta'$, and inter-node, $\eta \neq \eta'$, scatterings (corresponding to the C$_1$ and C$_2$ processes in Fig. 3C, respectively), whereas the $l$ th site Knight shift, $K_l$, probes only the intra-node ones (C$_1$). Employing the RG correction effects evaluated from fitting the Knight shift data at the site A in $a$-I$_3$ (hereafter A is not distinguished from A') (3), we have calculated the temperature dependence of $K_A$ and $(1/T_1T)_A$ for the non-interacting ($\alpha = 0$) and interacting ($\alpha \approx 8.4$) cases, as depicted in Fig. 3A. For each of these results, the Korringa ratio at the site A, $\mathcal{K}_A = (1/T_1TK^2)_A$, has been evaluated by assuming vertical cones (without a tilt; $\mathbf{w} = \mathbf{0}$) and tilted cones ($\mathbf{w} \neq \mathbf{0}$) (Fig. 3B). The calculated profiles of the corresponding susceptibilities in the first Brillouin zone are presented in Fig. 3, D and E, at high (25 K) and low (5 K) temperatures.

The contribution of the inter-node ($C_2$) fluctuations to the local spin susceptibility for the $l$ th orbital was calculated based on the ladder approximation as expressed by the following

$$\chi_{RL,ll}^{\perp}(\mathbf{q}, i\omega_m) = \sum_{\mathbf{k},\nu} [\mathcal{F}_{\nu\nu\bar{\nu}\bar{\nu}}^{RL,ll}\{\Lambda_+^\nu(\mathbf{k};\mathbf{q},i\omega_m)\chi_+^\nu(\mathbf{k};\mathbf{q},i\omega_m) + \Lambda_-^\nu(\mathbf{k};\mathbf{q},i\omega_m)\chi_-^\nu(\mathbf{k};\mathbf{q},i\omega_m)\}$$

$$+ \mathcal{F}_{\nu\bar{\nu}\nu\bar{\nu}}^{RL,ll}\{\Lambda_-^\nu(\mathbf{k};\mathbf{q},i\omega_m)\chi_+^\nu(\mathbf{k};\mathbf{q},i\omega_m) + \Lambda_+^\nu(\mathbf{k};\mathbf{q},i\omega_m)\chi_-^\nu(\mathbf{k};\mathbf{q},i\omega_m)\}],$$

(7)

where we defined

$$\chi_+^\nu(\mathbf{k};\mathbf{q},i\omega_m) = -T\sum_n G_R^{\nu\nu}(\mathbf{k}+\mathbf{q},i\varepsilon_n + i\omega_m)G_L^{\bar{\nu}\bar{\nu}}(\mathbf{k},i\varepsilon_n), \tag{8}$$

$$\chi_-^\nu(\mathbf{k};\mathbf{q},i\omega_m) = -T\sum_n G_R^{\nu\bar{\nu}}(\mathbf{k}+\mathbf{q},i\varepsilon_n + i\omega_m)G_L^{\nu\bar{\nu}}(\mathbf{k},i\varepsilon_n), \tag{9}$$

by means of the WF Green function $\hat{G}_\eta = [G_\eta^{\nu\nu'}]$, which takes into account the aforementioned RG correction effects (3). Note that the pseudospin base $\bar{\nu} = b(a)$ corresponds to $\nu = a(b)$ and $\varepsilon_n$ ($\omega_m$) is the fermionic (bosonic) Matsubara frequency. The ladder diagram (Fig. S2B) is given by

$$\Lambda_{\pm}^{\nu}(\mathbf{k};\mathbf{q},i\omega_m) = 1 + \sum_{\mathbf{k}'} V_0(\mathbf{k}-\mathbf{k}')[\Lambda_{\pm}^{\nu}(\mathbf{k};\mathbf{q},i\omega_m)\chi_{+}^{\nu}(\mathbf{k}';\mathbf{q},i\omega_m)$$
$$+\Lambda_{\mp}^{\nu}(\mathbf{k};\mathbf{q},i\omega_m)\chi_{-}^{\nu}(\mathbf{k}';\mathbf{q},i\omega_m)], \quad (10)$$

and

$$\mathcal{F}_{\nu_1,\nu_2,\nu_3,\nu_4}^{\eta,\eta',l,m} = d_{l\nu_1}(\eta\tilde{\mathbf{k}}_0')d_{m\nu_2}^*(\eta\tilde{\mathbf{k}}_0')d_{m\nu_3}(\eta'\tilde{\mathbf{k}}_0')d_{l\nu_4}^*(\eta'\tilde{\mathbf{k}}_0') \quad (11)$$

is the form factor expressed by the LK bases (*16*, *47*). Note that the filled area in the susceptibility diagram (Fig. S2A) corresponds to $\Lambda_{\pm}^{\nu}(\mathbf{k};\mathbf{q},i\omega_m)$. By separating the $\mathbf{k}$ dependence from the $(\mathbf{q},i\omega_m)$ dependence in Eq. 10, one yields $\Lambda_{\pm}^{\nu}(\mathbf{k};\mathbf{q},i\omega_m) \simeq \Delta_{\mathbf{k}}\Lambda_{\pm}^{\nu}(\mathbf{q},i\omega_m) \simeq (1-\lambda)^{-1}$ at the RPA level, where $\Delta_{\mathbf{k}}$ and $\lambda$ are given by

$$\lambda\Delta_{\mathbf{k}} = 2\sum_{\mathbf{k}'\nu} V_0(\mathbf{k}-\mathbf{k}')\Delta_{\mathbf{k}'}\{\chi_{+}^{\nu}(\mathbf{k};\mathbf{0},0) + \chi_{-}^{\nu}(\mathbf{k};0,0)\}. \quad (12)$$

The Eq. 12 is a linearized self-consistent equation that describes an excitonic instability favoring an opening of the gap $\Delta_{\mathbf{k}}$ at the node on $E_F$ when the eigenvalue $\lambda$ reaches unity (*30, 76*). For the excitonic phase transitions mediated by the long-range Coulomb interaction, extensive studies have been performed for the 2D WFs using various techniques such as mean field (*10, 11, 77-80*), Monte Carlo (*81*) and RG approaches (*82*). Similar instabilities have been also studied for the 3D WFs (*13, 83-90*).

To examine the excitonic instability (*10, 11, 30, 76-79, 81, 82*), the whole eight types of excitonic order parameters that generate a gap in 2D WF systems (*11, 91*) have been considered at a mean-field level, in which the RG correction effects owing to the long-range Coulomb interaction were incorporated. We find that the even-parity excitonic pairings in the inter-node ($C_2$) process, such as $\langle c_R^{a\dagger} c_L^b \rangle$, give dominant contributions towards reproducing the observed $1/T_1 T$ upturn (Fig. 4A). With decreasing temperature, the transverse spin susceptibility (Eq. 7), and thus $1/T_1 T$ (Eq. 3), increases because spin fluctuations develop through the enhancement of the ladder diagram as a precursor of excitonic instability, and in turn, diverges at the onset of the condensation (at $\lambda = 1$). The calculated temperature dependence of $1/T_1 T$ for different values of $\alpha$ is presented in Fig. 4B for the even-parity spin-triplet pairing, which shows an upturn for $\alpha = 2$ as a direct consequence of this precursor effect.

As to the eigenvalues of the linearized self-consistent equation (Eq. 12), we find that the even-parity, inter-node ($C_2$) and spin transverse instability (described by the eigenvalue $\lambda_\perp$) develops at low temperatures upon increasing the in-plane magnetic field $H$, whereas the spin longitudinal instability (given by the eigenvalue $\lambda_\parallel$) is suppressed with increasing $H$ (Fig. S3). Moreover, the intra-node ($C_1$) instability is perceived to be weak compared to the inter-node ($C_2$) instability (inset of Fig. S3). (See the Supplementary Text for details.)

It is worth noting that the titled cones in $\alpha$-I$_3$ are rather isotropic at low energy due to the RG flow (fig. S1) as mentioned above (*3*). This suggests that the excitonic spin fluctuations would directly manifest themselves in the relaxation rate $1/T_1 T$ (more sensitive to the transverse instability) even in the *isotropic* 2D WF systems such as graphene. In contrast, their influence would be limited on the spin shift $K$ that probes the longitudinal instability.

**Estimation of the critical temperature and the gap.** For the evaluation of the excitonic gap $\Delta$ and the critical temperature $T_C$, we employed Eqs. 1 and 2 together with the cutoff $\Lambda = 0.667$ Å$^{-1}$ of the size of the inverse lattice constant (*41*), the permittivity $\varepsilon \approx 30$ and the velocity $v = 2.4 \times 10^4$ m s$^{-1}$ determined from the fitting the Knight shift data reported previously (*3*).

**Supplementary Text:**

**Excitonic order parameters.** The WF states are described by an 8-component creation operator $\Psi_\mathbf{k}^\dagger = (c_{\mathbf{k},s,\eta}^\nu)^\dagger$, where $\nu$ ($= a, b$) represents the two pseudospin bases, $s$ ($= 1(\uparrow), -1(\downarrow)$) stands for the spin and $\eta$ ($= 1(R), -1(L)$) refers to the node. There are 64 possible order parameters, 8 of which can generate an excitonic mass gap. We can categorize these order parameters (*11, 91*) with respect to the total momentum referring to the intra-node and inter-node pairings (corresponding to the $C_1$ and $C_2$ processes in Fig. 3C), parity and spin. In Eqs. 10 and 12, for instance, even-parity pairings are assumed. Note that $\hat{s}_0$ ($\hat{s}_{x,y,z}$) corresponds to the spin-singlet (triplet) pairing. For the inter-node ($C_2$) case, we have even-parity operators:

$$\Phi^{2\mathbf{k}_0,e}(\mathbf{k};\mathbf{q}) = \Psi_{\mathbf{k}+\mathbf{q}}^\dagger (\hat{\sigma}_x \otimes \hat{\tau}_x \otimes \hat{s}_t)\Psi_\mathbf{k}$$

$$= \left(c_{\mathbf{k}+\mathbf{q},s,R}^{a\dagger} c_{\mathbf{k},s',L}^{b} + c_{\mathbf{k}+\mathbf{q},s,R}^{b\dagger} c_{\mathbf{k},s',L}^{a} + c_{\mathbf{k}+\mathbf{q},s,L}^{b\dagger} c_{\mathbf{k},s',R}^{a} + c_{\mathbf{k}+\mathbf{q},s,L}^{a\dagger} c_{\mathbf{k},s',R}^{b}\right)[\hat{s}_t]_{ss'},$$

(13)

where **k** and **k** $\pm$ **q** are restricted to the vicinity of the same node. (The wavenumber vector **q** is assumed to be much smaller compared to the size of the nesting vector $2\tilde{\mathbf{k}}_0$ connecting the two nodes at $\tilde{\mathbf{k}}_0$ and $-\tilde{\mathbf{k}}_0$ in the first Brillouin zone.) In general, these order parameters give contributions to both the density waves and bond order waves, since the pseudospin bases (corresponding to $\hat{\sigma}_i$) are not always described by the sublattice bases as in graphene but can be generally expressed by the LK pseudospin bases (*67*) such as for the 2D tilted WFs in $\alpha$-I$_3$ (*17, 18*).

**The linearized gap equation for the excitonic instabilities.** In Fig. S3, we show the temperature dependence of the eigenvalue $\lambda$ of the linearized gap equation (Eq. 12) for selected values of the in-plane magnetic field $H$, where $\lambda$ reaches unity at the onset of excitonic condensation. The calculations are performed for the even-parity, spin-triplet excitonic pairings, where we assumed the dimensionless Coulomb coupling of $\alpha = 1$ and considered the velocity renormalization due to the self-energy correction (*26*). The inset shows the eigenvalues for the intra-node ($C_1$ in Fig. 3C) and inter-node ($C_2$) pairings at $H = 0$ T. Because the electron-hole symmetry is higher for the $C_2$ excitonic pairing than for the $C_1$ pairing, the excitonic instability in the $C_2$ process develops stronger upon cooling. With increasing $H$, the eigenvalue for the transverse instability $\lambda_\perp$ is enhanced, whereas that for the longitudinal instability $\lambda_\parallel$ is suppressed. This is related to the fact that the Weyl cone in $\alpha$-I$_3$ is tilted, and the Zeeman-induced Fermi pockets have an elliptical shape (*16-18, 22, 37, 45-47*). Hence, there is a perfect nesting between the up-spin and down-spin pockets, while the nesting between the same-spin pockets is poor, causing the different natures in the transverse and longitudinal instabilities.

The odd-parity excitonic orders corresponding to the topological Mott insulators or the flux states are expected when the next-nearest neighbor repulsions are larger than the nearest neighbor ones (*80*). In the present study, however, such exotic orders do not realize since the Coulomb potential ($\propto 1/|\mathbf{r}|$) is considered.

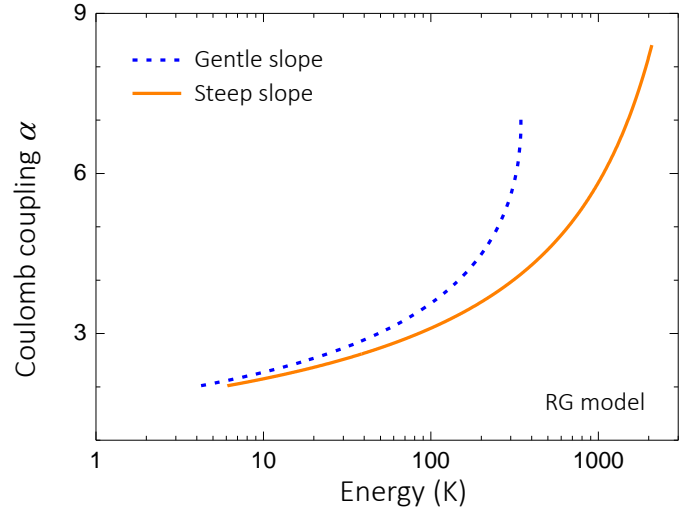

**Fig. S1. Anisotropic RG flow of the Coulomb coupling in $\alpha$-(BEDT-TTF)$_2$I$_3$.** Parameters used for the calculation are $\varepsilon \approx 30$ and the velocity $v = 2.4\times10^4$ m s$^{-1}$ determined from fitting the Knight shift data using RG analyses, whose details are described in Ref. *3* (*26*). The dimensionless Coulomb coupling $\alpha$, which is highly anisotropic around the Weyl node reflecting the tilt of the Weyl cone in $\alpha$-(BEDT-TTF)$_2$I$_3$, is evaluated at the gentle slope (dashed trace) and the steep slope (solid trace) of the tilted cone.

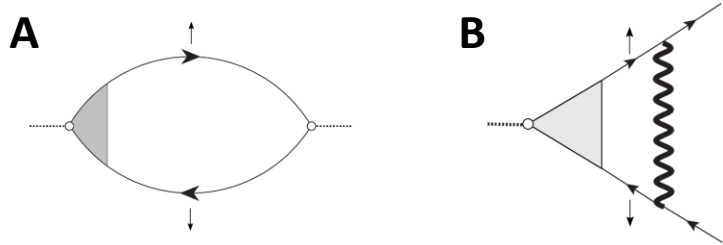

**Fig. S2. Spin susceptibility and Ladder diagram.** (**A** and **B**) **The** Feynman diagrams for the spin susceptibility (Eq. 7) (A) and the ladder diagram $\Lambda^\nu_+(\mathbf{k}; \mathbf{q}, i\omega)$ (Eq. 10) (B). The renormalized fermion propagator due to the self-energy effect (solid line) and the RPA boson propagator (wavy line) are incorporated. The up and down arrows indicate corresponding spin projections.

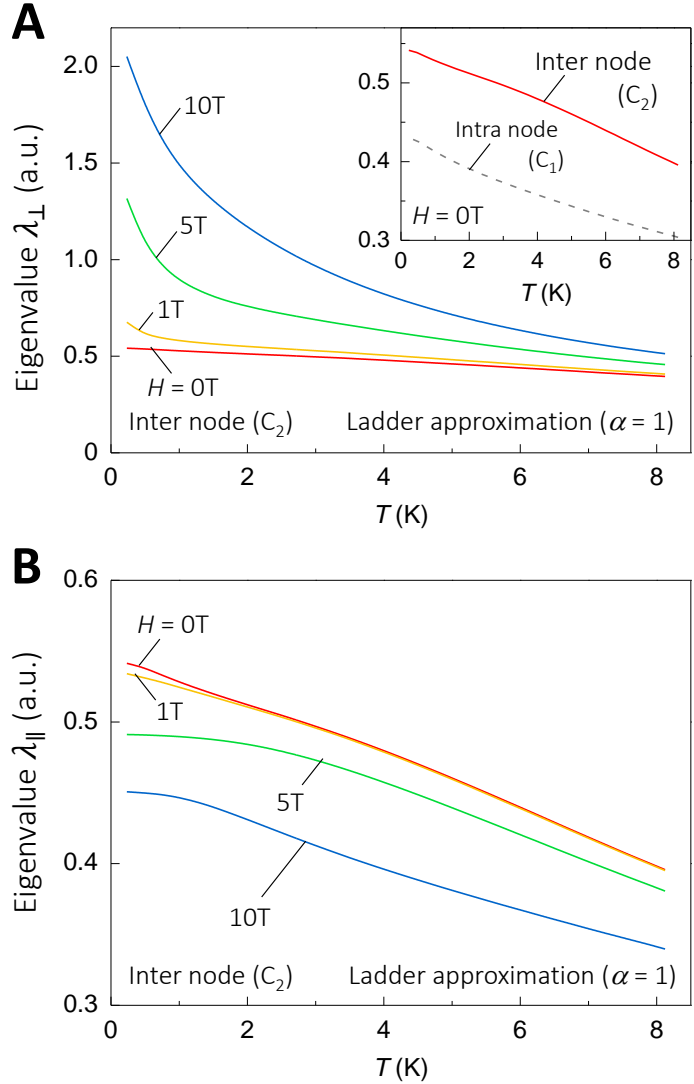

**Fig. S3. Eigenvalues for the excitonic pairings.** (**A** and **B**) The temperature dependence of the eigenvalue of the linearized gap equation (Eq. 12) for the even-parity spin-triplet excitonic pairings derived based on the ladder approximation, with the velocity renormalization taken into account (*26*). The dimensionless Coulomb coupling of $\alpha = 1$ is employed in the calculation. The eigenvalues for the spin transverse instability $\lambda_\perp$ (A) and the spin longitudinal instability $\lambda_\parallel$ (B) for the inter-node pairing ($C_2$ in Fig. 3C) are plotted, calculated for the in-plane magnetic field $H$ of 0 T (red), 1 T (orange), 5 T (green) and 10

T (blue). At $H = 0$ T, $\lambda_\perp$ and $\lambda_\parallel$ are degenerate, while $\lambda_\perp$ grows towards lower temperature much stronger than $\lambda_\parallel$ at a finite field. The eigenvalues for the intra-node ($C_1$ in Fig. 3C) and inter-node ($C_2$) pairings at $H = 0$ are given in the inset of (A).